\documentclass[a4paper,11pt]{article}
\pdfoutput=1 

\usepackage{jheppub} 

\usepackage[T1]{fontenc} 

\usepackage{braket}
\usepackage{todonotes}
\usepackage{multirow}
\usepackage{subfiles}
\usepackage{circuitikz}
\usepackage{tikz-cd}

\newcommand{\Nhat}{\hat{N}}
\newcommand{\Shat}{\hat{S}}
\newcommand{\Lc}{\mathcal{L}}
\newcommand{\ZZ}{\mathbb{Z}}
\newcommand{\chihat}{\hat{\chi}}
\newcommand{\chibar}{\bar{\chi}}

\newcommand{\chichi}[2]{\chi_{#1}\bar{\chi}_{#2}}
\newcommand{\chichihat}[2]{\chihat_{#1}\bar{\chihat}_{#2}}

\newenvironment{psmallmatrix}{\left(\begin{smallmatrix}}{\end{smallmatrix}\right)}
\newcommand{\Zbox}[2]{{\scriptstyle #1}\underset{#2}{\square}}

\title{\boldmath Topological Defect Lines in bosonized Parafermionic CFTs}

\author[\ast,\dag]{Babak Haghighat}
\author[\ast]{and Sun Youran}
\affiliation[\ast]{Yau Mathematical Sciences Center, Tsinghua University, Beijing, 100084, China}
\affiliation[\dag]{Yanqi Lake Beijing Institute of Mathematical Sciences and Applications (BIMSA), Huairou District, Beijing 101408, China}

\emailAdd{babakhaghighat@tsinghua.edu.cn}
\emailAdd{syouran0508@gmail.com}

\abstract{Topological defect lines (TDLs) are extended line operators which act on the Hilbert space of two-dimensional CFTs and satisfy non-trivial fusion algebras when forming junctions. Among the most interesting fusion algebras are the so-called Tambara-Yamagami (TY) fusion categories which are realized in (bosonized) Parafermionic CFTs. The corresponding TY[$\mathbb{Z}_k$]-categories have been explicitly realized for the cases $k=2$, $k=3$, and $k=4$ together with the action of the defect lines on the Hilbert space of the corresponding CFTs. For each of the cases, different methods have been used in the previous literature. In the current paper, we present a unified framework for finding the TDLs in bosonized Parafermionic CFTs. Our approach relies on generalizing several previously used methods by introducing the notion of an extended $S$ matrix. We apply the method to the cases $k=2$ to $k=5$ to extract corresponding TDL fusion algebras.}

\begin{document} 
\maketitle
\flushbottom

\section{Introduction}

Topological defects are associated with global symmetries which act on point-like or higher-dimensional operators.
In the former case, they generate usual global symmetries while in the latter case, they become generators of so-called \textit{higher-form} symmetries \cite{Kapustin:2013uxa, Gaiotto:2014kfa}.
Recently, this viewpoint has led to a generalization of the concept of symmetry \cite{ Morrison:2020ool, Albertini:2020mdx, Apruzzi:2020zot, Braun:2021sex, Closset:2020scj, Bhardwaj:2020phs, Closset:2020afy, Apruzzi:2021phx, Hosseini:2021ged,  Bhardwaj:2021mzl, Buican:2021xhs, Cvetic:2021maf, Cvetic:2021sxm, Cvetic:2021sjm, Closset:2021lwy, Lin:2021aps, Genolini:2022mpi, DelZotto:2022fnw, Cvetic:2022uuu}, including non-invertible symmetries
\cite{Verlinde:1988sn, Chang:2018iay, Thorngren:2019iar, Thorngren:2021yso, Choi:2022zal, Chen:2023qnv, Schafer-Nameki:2023jdn}
which shall be of relevance in the current work.

On the other hand, in the framework of 2d CFTs, extended defects have been known for quite some while now, and have been realized through a choice of boundary conditions \cite{Cardy:1986gw, Verlinde:1988sn, Cardy:1989, Cardy:2004hm} or as line defects/interfaces \cite{Oshikawa:1996ww, Oshikawa:1996dj, Petkova:2000ip, Fuchs:2002cm, Fuchs:2003id, Fuchs:2004dz, Fuchs:2004xi, Frohlich:2004ef, Frohlich:2006ch, Quella:2006de, Fuchs:2007vk, Fuchs:2007tx, Bachas:2007td,  Kong:2009inh, Petkova:2009pe, Carqueville:2012dk, Brunner:2013xna, Davydov:2013lma, Kong:2013gca, Petkova:2013yoa, Gu:2021utd, Gu:2023bjg}
or in machine learning applications \cite{Noormandipour:2021}. 
Apart from the usual Verlinde defects corresponding to primary operators, 2d CFTs can admit other types of defects including Kramers-Wannier duality defects\footnote{For the Ising model, the Kramers-Wannier duality defect happens to be the Verlinde defect associated to $\sigma$ field.}.
The corresponding fusion categories are known as the Tambara-Yamagami category $\mathrm{TY}[\mathbb{Z}_k]$ \cite{Tambara1998} where $k$ is a positive integer greater or equal to $2$. The most famous example is the Ising model, which admits a duality interface upon inverting the coupling constant.
At the self-critical conformal point, this interface becomes a defect with non-trivial fusion properties. 
In particular, when fused to itself the result is the sum of the transparent defect and the spin-flip defect corresponding to the fermion field $\epsilon$ which together form a $\mathbb{Z}_2$ Abelian fusion category. The natural extension to higher $k$ arises in parafermionic CFTs \cite{Fateev:1985mm} which can be understood as cosets of WZW models \cite{Gepner:1986hr, Mathieu2000, Mathieu:parafermioncharacter} after a bosonization procedure described in \cite{Thorngren:2021yso}. It is this bosonized version of parafermionic CFTs we are interested in, and for notational simplicity, we henceforth denote them by $\mathbb{Z}_k$ Parafermionic CFTs. For $k=3$ one obtains the $\mathbb{Z}_3$ Parafermionic CFT which is closely related to the 3-state Potts model. The corresponding defect lines were obtained in recent literature from two different perspectives, namely from a boundary conformal field theory perspective in \cite{Petkova:2000ip} and from a fusion category perspective in \cite{Chang:2018iay}. For $k=4$, the corresponding CFT has central charge $c=1$ and is related to the compact boson CFT \cite{Thorngren:2021yso} where also the action of the duality defect on the Hilbert space is worked out. The next interesting example is the case $k=6$, where a supersymmetry current can be realized \cite{Bae:2021lvk}. All these examples contain the TY category as a subcategory of defects and are highly relevant for condensed matter applications \cite{Aasen:2020jwb, Mong:2014ova}. We find that the case of $k=4$ is rather subtle as our method can only correctly reproduce the fusion rules of the Verlinde and TY subcategories of the entire fusion category. 

The goal of our current work is to generalize the approaches of \cite{Chang:2018iay} and \cite{Petkova:2000ip} to streamline the formalism for extracting topological defect lines and their fusion rules. 
To this end, we introduce the notion of an \textit{extended S-matrix} $\hat{S}$ which can be used to derive a generalized version of the Verlinde formula. 
As examples, we will treat the Parafermionic coset CFTs with $k=2$, $k=3$, $k=4$, and $k=5$, where the truly non-trivial cases are the last three. 
In these cases, the action of the duality defect on the Hilbert space, when wrapped around the spacial circle, is not diagonalized by the standard basis of Parafermionic characters. 
Thus we propose to split some of these characteristics into a two-dimensional state space where the duality defect acts diagonally. 
This is where $\hat{S}$ comes into play and we use residual basis transformations to render
this new matrix unitary. 
One can then show that a generalized Verlinde formula \eqref{eq:generalVerlinde} holds which involves all defect fusion symbols and not only those of classical Verlinde lines.
Summarizing, we obtain the following diagram
\begin{equation}
\begin{tikzcd}[row sep=large, column sep=large]
S\textrm{ matrix}
\arrow{rrr}{\textrm{split some characters}} 
\arrow[swap]{d}
&&& \textrm{extended }S\textrm{ matrix }\hat{S}
\arrow{d} \\
\textrm{fusion algebra }N \arrow[hook]{rrr}
&&& \textrm{enlarged fusion algebra }\hat{N}~.
\end{tikzcd} \nonumber
\end{equation}

Consistency of the whole approach requires that the resulting defect fusion symbols satisfy an associativity property and hence obey the defect fusion algebra. We remark that for the entire approach, including the construction of the extended $S$ matrix, the actual knowledge about how characters split into two new characters is not necessary. This is best exemplified in the $k=5$ case, where the splitting is not known, yet self-consistency fixes the extended S-matrix and with it the defect fusion rules. 

The organization of the present paper is as follows. In Section \ref{sec:TDL}, we review the notion of topological defect lines, their properties, and the construction of corresponding partition functions. In Section \ref{sec:parafermion}, we summarize the main properties of $\mathbb{Z}_k$ Parafermionic coset CFTs and write down explicit formulas for their characters. We also point out in the example of the $k=3$ case how characters split when comparing the Parafermionic theory with the 3-state Potts model. This will provide important lessons for generalizations later on. In Section \ref{sec:find-tdl} we outline our main approach to finding topological defect lines by making use of boundary fusion symbols and the S-transformation. We then proceed to apply the method explicitly to the cases of $k=2$ to $k=5$. Finally, in Section \ref{sec:conclusions} we present our Conclusions. 

\section{Topological Defect Lines}
\label{sec:TDL}

Traditionally, symmetries have been described as invertible operations on the Hilbert space.
However, recent understanding indicates that symmetries in QFT are linked to topological operators.
For instance, a symmetry that induces field transformations can be represented by an invertible topological defect surface.
By setting the boundary condition across this surface, the fields on one side can be connected to the transformed fields on the other side.
This operator encapsulates the entirety of the symmetry, leading us to define symmetry as an invertible topological operator with codimension one \cite{Thorngren:2021yso}.

One generalization is to relax the condition of invertibility to the definition of topological defect line (TDL).
A topological defect line $X$ is a line operator acting on the Hilbert space of a CFT.
It commutes with the energy-momentum tensor $T(z)$, $\overline{T}(\bar{z})$, or equivalently with the Virasoro generators
\begin{equation}\label{eq:[Ln,X]}
    [L_n,X]=[\bar{L}_n,X]=0.
\end{equation}
Given that the Virasoro operators are responsible for generating local conformal transformations, this condition implies that the precise position and shape of the line to which $X$ is attached are irrelevant. In other words, the defect lines exhibit topological properties.

The primaries of a CFT constitute a fusion category.
Likewise, the topological defect lines (TDLs) of a CFT also constitute a fusion category. 
This means that TDLs can combine and separate into other TDLs, forming interconnected networks.
The fusion rule of TDLs may or may not be identical to the fusion rule of primaries of the CFT.
The property of topological invariance implies that these networks can be altered through homotopy, away from or approaching any inserted operators, without affecting any correlation function with the background of such a network.

\subsection{Fusion Category}\label{sec:fusioncategory}

\begin{figure}[ht]
    \centering
    \includegraphics[width=\textwidth]{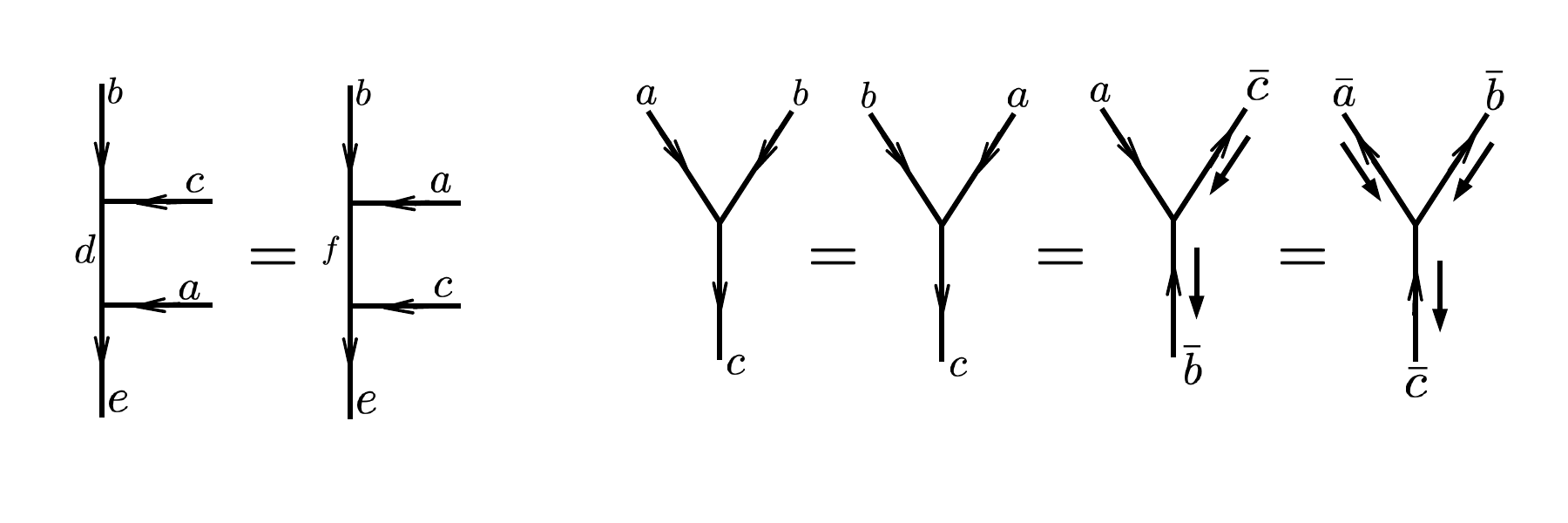}
    \caption{Illustration of fusion symbols' properties. The left figure is for Eq. \eqref{eq:NN=NN}, while the right figure for Eq. \eqref{eq:Nsym}.}
    \label{fig:Nproperty}
\end{figure}

A fusion category \cite{Etingof_2005} comprises a finite number of \textit{simple objects} denoted as $a,b,c..$.
The fusion rules are specified by non-negative integers $N^c_{ab}$ which dictate the fusion of two simple objects labeled by $a$ and $b$ into $c$
\begin{equation}
    a\otimes b=\bigoplus_c N_{ab}^c\; c.
\end{equation}
The terms $N_{**}^*$ are referred to as \textit{fusion symbols}.
Within the collection of simple objects, there exists a special \textit{identity} object labeled as 1.
For any object $a$, the fusion of 1 with $a$ or $a$ with 1 results in $a$ itself: $1\otimes a=a\otimes 1=a$. Consequently, this implies that $N_{a1}^b=N_{1a}^b=\delta_{ab}$.
For every object $a$ there is a unique \textit{dual} object denoted as $\bar{a}$, where the fusion of $a$ and $\bar{a}$ contains the identity object: $a\otimes\bar{a}=1\oplus..$.
If a simple object is identical to its dual, it is referred to as \textit{self-dual}.
The fusion rules within a fusion category must satisfy the associativity property
\begin{equation}\label{eq:NN=NN}
    a\otimes(b\otimes c)=(a\otimes b)\otimes c \quad\Rightarrow\quad
    \sum_{d}N_{bc}^dN_{ad}^e=\sum_{f}N_{ab}^fN_{cf}^e.
\end{equation}
What's more, the fusion symbols obey
\begin{equation}\label{eq:Nsym}
N_{ab}^c=N_{ba}^c=N_{a\bar{c}}^{\bar{b}}=N_{\bar{a}\bar{b}}^{\bar{c}}.
\end{equation}
The above constraints can be simply visualized by fusion graphs as in Figure \ref{fig:Nproperty}.

A familiar example of fusion rules comes from taking irreducible representations $A$, $B$, $C...$ of a finite group as simple objects and tensor products of representations as fusion.
The fusion rules specify how to write the tensor product $A\otimes B$ as sums over irreps.
A physical example arises when simple objects are taken to be the primaries of a CFT, while the fusion rules are their OPE, i.e. $N_{ab}^c$ counts the number of times the primary $c$ appears in the OPE of $a$ and $b$.
In our work, the topological defect lines of a CFT form a fusion category.

Any correlation function evaluated in the background of a TDL network is unaffected by the topological movements of the TDLs.
To evaluate a TDL network, one needs further data, which are given by the \textit{$F$-symbols}.
The $F$-symbols are defined as
\begin{equation}
\def \yabc {8.6}
\def \yd   {4.4}
\def \yshift {-0.5ex}
\begin{tikzpicture}[baseline={([yshift=\yshift]current bounding box.center)},scale=0.4]
    \node at (-5,\yd) {$d$};
    \node at (-7,\yabc) {$a$};
    \node at (-5,\yabc) {$b$};
    \node at (-3,\yabc) {$c$};
    \draw (-5,5) -- (-5,6);
    \draw (-5,6) -- (-6,7) node[below] {$x\ $};
    \draw (-5,6) -- (-4,7);
    \draw (-6,7) -- (-7,8);
    \draw (-6,7) -- (-5,8);
    \draw (-4,7) -- (-3,8);
\end{tikzpicture}
=\sum_y \left(F_d^{abc}\right)_{xy}
\begin{tikzpicture}[baseline={([yshift=\yshift]current bounding box.center)},scale=0.4]
    \node at (5,\yd) {$d$};
    \node at (7,\yabc) {$c$};
    \node at (5,\yabc) {$b$};
    \node at (3,\yabc) {$a$};
    \draw (5,5) -- (5,6);
    \draw (5,6) -- (6,7) node[below] {$\ y$};
    \draw (5,6) -- (4,7);
    \draw (6,7) -- (7,8);
    \draw (6,7) -- (5,8);
    \draw (4,7) -- (3,8);
\end{tikzpicture}
\end{equation}
They describe linear relations among diagrams and provide the means to evaluate an arbitrarily labeled planar diagram.
Different ways of simplifying a graph by $F$-moves give different evaluations of the same graph.
However, they all come from the same graph, so they must be equal.
This gives strong constraints on the $F$-symbols such as the \textit{pentagon identity}.
Two fusion categories with the same fusion rule may have different $F$-symbols and so be considered as different fusion categories.
For a more detailed overview, we refer to \cite{Aasen:2020jwb}.

\subsection{Partition functions with defect lines}\label{sec:tdl}

Consider a CFT and denote by  $\mathcal{I}$ its primaries, by $\chi_i(q)$, $S_{ij}$, $N_{ij}^k$ with $i,j,k\in \mathcal{I}$ their characters, modular $S$ matrix and fusion rule.
In the rest of this paper, we will use $i,j,k...$ for primary field indices and $a,b,c...$ for indices of TDLs.
It is a common practice to consider the CFT on a cylinder.
If the two boundaries of this cylinder are identified by periodic boundary conditions, then the CFT is regarded as living on a torus and the partition function is the trace of the time evolution operator.
However, we shall allow the insertion of one operator $X$ inside the trace.
This can be interpreted as introducing a twisted boundary condition or introducing a defect line along one non-contractible cycle of the cylinder. 
For the defect line to be topological, it should commute with the Virasoro generators as shown in Eq. \eqref{eq:[Ln,X]}.

\begin{figure}
    \centering
    \includegraphics[width=.6\textwidth]{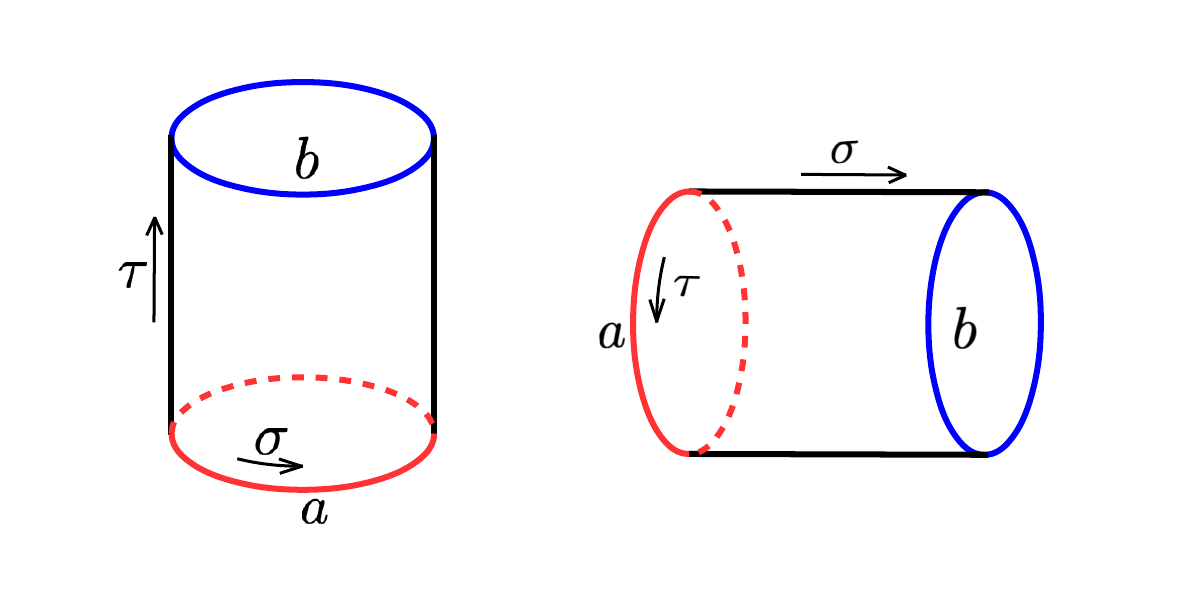}
    \caption{Two ways to compute $Z_{a|b}$. On the left: periodic in space, between states $a$ and $b$; on the right: periodic in time, between boundary condition $a$ and $b$.}
    \label{fig:Zab}
\end{figure}

Following Cardy \cite{Cardy:2004hm}, consider the partition function 
\begin{equation}
    Z_{b|a}:=\bra{b}q^{\frac{1}{2}(L_0+\bar{L}_0-\frac{c}{12})}\ket{a},
\end{equation}
where $\ket{b}$ and $\ket{a}$ are boundary states satisfying
\begin{equation}\label{eq:res-on-a}
    (L_n+\bar{L}_n)\ket{a}=0,
\end{equation}
as shown in Figure \ref{fig:Zab}.
The solutions to Eq. \eqref{eq:res-on-a} are known as Ishibashi basis states
\begin{equation}
    \ket{a}=\sum_{j}\frac{\psi^j_{a}}{\sqrt{S_{1j}}}\ket{j},
\end{equation}
in which the $\sqrt{S_{1j}}$ is for later convenience.
Viewing $Z_{b|a}$ as periodic in space and evolution in time, we can see that
\begin{equation}
    Z_{b|a}=\sum_{j}\frac{\psi^j_a(\psi^j_{b})^*}{S_{1j}}\chi_{j}(-1/\tau)
\end{equation}
where $\tau$ is related to $q$ by $q=e^{2\pi i \tau}$.
On the other hand, if we act by an $S$-transformation on $Z_{b|a}$ and view the result as periodic in time, then it should be a linear composition of characters
\begin{equation}\label{eq:Zba-niba}
    Z_{b|a}=\sum_{i}n_{ib}^a\chi_{i}(\tau).
\end{equation}
Comparing the above two formulas, we get an expression for $n_{ib}^a$~,
\begin{equation}
    n_{ib}^a=\sum_{j}\frac{S_{ij}}{S_{1j}}\psi^j_{b}(\psi^j_a)^*.
\end{equation}
Assuming that the Ishibashi basis is orthogonal and complete, i.e., $\sum_{j}\psi^j_a(\psi^j_{b})^*=\delta_{ab}$ and $\sum_{a}\psi^i_a(\psi^j_{a})^*=\delta_{ij}$, and using the Verlinda formula, 
\begin{equation}\label{eq:Verlinde-formula}
    \frac{S_{i_1j}S_{i_2j}}{S_{1j}S_{1j}}=N_{i_1i_2}^{i_3}\frac{S_{i_3j}}{S_{1j}},
\end{equation}
one finds that $n_{1a}^b$ forms a representation of primary operator fusion rules,
\begin{equation}\label{eq:nn=Nn}
    n_{i_1}n_{i_2}=\sum_{i_3}N_{i_1i_2}^{i_3}n_{i_3}.
\end{equation}
Then we define the \textit{defect line fusion rules} as 
\begin{equation}
    N_{ab}^c:=\sum_{j}\frac{\psi^j_a\psi^j_b(\psi^j_c)^*}{\psi^j_1}.
\end{equation}
$N_{ab}^c$ are non-negative integers for RCFTs with a block diagonal partition function.
Parafermion theories fall into this case.
From the definition, we can see that
\begin{equation}\label{eq:n=nN}
    n_{ia}^b=n_{i1}^c N_{ca}^b.
\end{equation}
By introducing the \textit{block characters}
\begin{equation}
    \chi_b:=n_{i1}^b\chi_i,
\end{equation}
and using Eq. \eqref{eq:n=nN}, Eq. \eqref{eq:Zba-niba} can be rewritten as
\begin{equation}
    Z_{b|a}=N_{cb}^a\chi_{c}.
\end{equation}
In other words, the defect line $\Lc_b$ acts on $\chihat_a$ giving
\begin{equation}\label{eq:Lchi}
\mathcal{L}_b\chi_{a}=\sum_{c}N_{ab}^c\chi_{c}.
\end{equation}
This gives an intuitive definition of TDLs: namely that they are topological interfaces that connect two regions with different boundary conditions.

\section[The Zk Parafermionic CFTs]{The $\mathbb{Z}_k$ Parafermionic CFTs}\label{sec:parafermion}

The $\mathbb{Z}_k$ Parafermionic CFT is a natural generalization of the Ising model in which the Ising model can be viewed as $\mathbb{Z}_2$ Parafermionic.
It can be constructed as a coset of WZW models
\begin{equation}
    \frac{\hat{su}(2)_k}{\hat{u}(1)}.
\end{equation}
Its central charge is
\begin{equation}
    c=c[\hat{su}(2)_k]-c[\hat{u}(1)]=\frac{2(k-1)}{k+2}.
\end{equation}
The central charges for $k$ ranging from 2 to 8 are shown in Table \ref{tab:para-c}.
The primaries of Parafermion theories are labeled by two integers: $l=1,\cdot\cdot\cdot,k$ and $-k<m\leq k$ s.t. $m+l$ even. 
The number of primaries is $k(k+1)/2$.
Their conformal weights are
\begin{equation}
    h_{l,m}=\frac{l(l+2)}{4(k+2)}-\frac{m^2}{4k}~.
\end{equation}
The conformal weights for $\mathbb{Z}_3$, $\mathbb{Z}_4$ and $\mathbb{Z}_5$ parafermions are shown in Tables \ref{tab:wt-z3}, \ref{tab:wt-z4} and \ref{tab:wt-z5}.
There are many equivalent formulas for characters for the $\mathbb{Z}_k$ parafermionic theory, here is the historically earliest one \cite{Mathieu:parafermioncharacter},
\begin{equation}\begin{aligned}
    \chi_{l,m}=\frac{1}{\eta^2(q)}
    &\left\{\left(\sum_{i\geq0,j\geq0}-\sum_{i<0,j<0}\right)(-1)^iq^{\frac{(l+1+(i+2j)(k+2))^2}{4(k+2)}-\frac{(m+ik)^2}{4k}}\right.\\
    &-\left.\left(\sum_{i\geq0,j>0}-\sum_{i<0,j\leq0}\right)(-1)^iq^{\frac{(l+1-(i+2j)(k+2))^2}{4(k+2)}-\frac{(m+ik)^2}{4k}}\right\}.
\end{aligned}\end{equation}
The $\mathcal{S}$ matrix for the Parafermion theory is
\begin{equation}\label{eq:pafaS}
    \mathcal{S}_{l,m;l',m'}=\frac{2}{\sqrt{k(k+2)}}e^{2\pi i\frac{mm'}{2k}}\sin\left(\pi\frac{(l+1)(l'+1)}{k+2}\right).
\end{equation}

\begin{table}[ht]
    \centering
    \begin{tabular}{c|ccccccc}
$k$& 2& 3& 4& 5 & 6& 7& 8\\
\hline
Central charge $c$& $\frac{1}{2}$ & $\frac{4}{5}$& 1& $\frac{8}{7}$ &$\frac{5}{4}$& $\frac{4}{3}$&$\frac{7}{5}$
    \end{tabular}
    \caption{Central charges for $k=2,3,4,5,6,7,8$ $\ZZ_k$ Parafermion theories.}
    \label{tab:para-c}
\end{table}

\begin{table}[ht]
    \centering
    \begin{tabular}{c|ccccc}
        $m$&-1&0&1&2&3\\
        \hline
        $l=1$&&& $\frac{1}{15}$\\
        $l=2$&& $\frac{2}{5}$&& $\frac{1}{15}$\\
        $l=3$& $\frac{2}{3}$&& $\frac{2}{3}$&& $0$
    \end{tabular}
    \caption{Conformal weights for $\ZZ_3$ Parafermion primaries.}
    \label{tab:wt-z3}
\end{table}


\subsection[The relationship between Z3 Parafermion and 3-state Potts model]{The relationship between $\ZZ_3$ Parafermions and the 3-state Potts model}

One coincidence we want to point out now, and which will be useful later on, is the relation between the $\ZZ_3$ Parafermion theory and the 3-state Potts model.
Both of them have a central charge of $4/5$.
By comparing the characters $\chi_{l,m}$ of the $\ZZ_3$ Parafermionic CFT with those of the 3-state Potts model, one can confirm that
\begin{equation}\label{eq:Z3-Potts-char}\begin{aligned}
    \chi_{3,3}&=\chi_{h=0}+\chi_{h=3},\quad
    \chi_{2,0}=\chi_{h=2/5}+\chi_{h=7/5}\\
    \chi_{1,1}&=\chi_{2,2}=\chi_{h=1/15},\quad
    \chi_{3,-1}=\chi_{3,1}=\chi_{h=2/3}
\end{aligned}\end{equation}
This agrees with Eq. (3.12) of \cite{Mathieu2000}.
The $\ZZ_3$ Parafermion has 6 primaries but only 4 different characters, in which $\chi_{1,1}=\chi_{2,2}$ and $\chi_{3,-1}=\chi_{3,1}$.
The fields in the 3-state Potts model have some traditional symbols as shown in Table \ref{tab:Potts-alias}.
\begin{table}[ht]
    \centering
    \begin{tabular}{c|cccccc}
    Conformal Dimension & 0 & 3 & $\frac{2}{5}$ & $\frac{7}{5}$ & $\frac{2}{3}$ & $\frac{1}{15}$\\[0.5ex]
    \hline
    Symbol & $I$ & $Y$ & $\epsilon$ & $X$ & $Z$ & $\sigma$
    \end{tabular}
    \caption{Traditional symbols for 3-state Potts model primaries}
    \label{tab:Potts-alias}
\end{table}

\subsection{Duality Defects}
\label{sec:fermionization}

A bosonic CFT is dual by gauging/fermionization to another bosonic/fermionic CFT \cite{Ji_2020,Lawrie_2024}. Although the fermionization operation is not relevant to the current paper, we briefly mention it here. Starting from a bosonic CFT with a non-anomalous $\ZZ_2$ symmetry generated by $g$, one can fermionize it by first stacking a Kitaev chain and then modding out the $\ZZ_2$ symmetry. The partition functions are related by
\begin{equation}
    Z_F[\alpha]=\frac{1}{2}\sum_{\rho}(-1)^{\textrm{Arf}[\rho+\alpha]}Z_{B}[\rho],
\end{equation}
where the $\alpha$ and $\rho$ are spin structures and $\textrm{Arf}$ is the Arf invariant.
Practically, it is computed on a torus following the rule
\begin{equation}
    \textrm{Arf}[\rho]=\begin{cases}
    1\quad\textrm{if}\;\rho=\Zbox{g}{g}\\
    0\quad\textrm{otherwise.}
    \end{cases}
\end{equation}
The more interesting operation for us is the gauging operation for the non-anomalous $\mathbb{Z}_k$ symmetries of bosonized Parafermionic CFTs. For the case of torus partition functions, the corresponding gauging operation, which gives rise to the partition function of a dual theory $B'$, is given as follows
\begin{equation}
    Z_{B'}[b] = \frac{1}{k} \sum_{a_1, a_2 \in \mathbb{Z}_k} \omega^{a_1 b_2 - b_1 a_2} Z_B[a], 
\end{equation}
where $\omega = e^{\frac{2\pi i}{k}}$ and $b=(b_1,b_2)$, $a=(a_1,a_2)$ are the (background) gauge field expectation values along the two cycles of the torus. For Parafermionic CFTs, this gauging operation is self-dual. That is the bosonic theory $B$ and the dual bosonic theory $B'$ are the same and their characters just get permuted by the gauging operation. This gives rise to a topological defect line in such theories connecting two regions corresponding to the original and the gauged theory. 
This defect actually gives rise to the duality defect $N$ of a TY[$\mathbb{Z}_k$] fusion category as explained in \cite{Thorngren:2021yso}. A TY[$A$] category consists of objects $L_a,L_b,...$ corresponding to an Abelian group $A$ and a single non-invertible object $N$.
Their fusion rule is
\begin{equation}
    L_a\otimes L_b=L_{(a+b)},\quad L_a\otimes N=N,\quad N\otimes N=\bigoplus_{a\in A}L_a.
\end{equation}


\section{Finding Topological Defect Lines}\label{sec:find-tdl}

\subsection{Basic Setup}\label{sec:basicsetups}

\begin{figure}[ht]
    \centering
    \includegraphics[width=.25\textwidth]{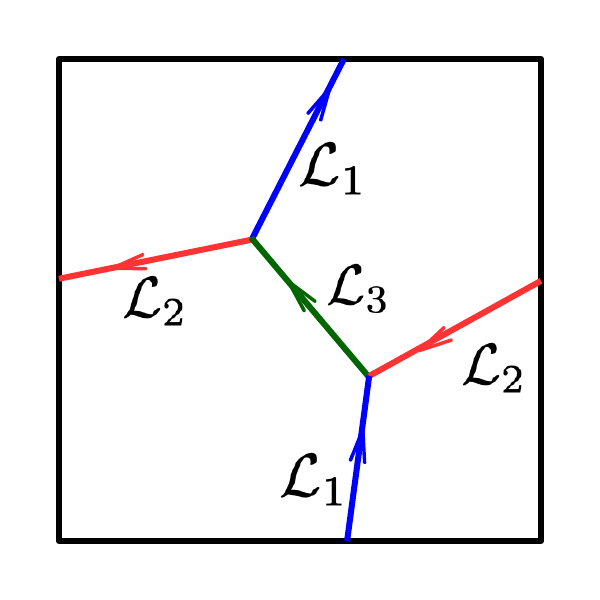}
    \caption{A torus with $\mathcal{L}_1$ inserted in the time direction and $\mathcal{L}_2$ inserted in the space direction}
    \label{fig:Zl1l2l3}
\end{figure}

We are interested in the partition function of a TDL network on a torus. Figure \ref{fig:Zl1l2l3} shows a torus with $\Lc_1$ inserted in the time direction and $\Lc_2$ inserted in the space direction.
Its partition function is denoted as $Z_{\Lc_1\Lc_2}^{\Lc_3}$.
When $\Lc_1$ is trivial, i.e. $\Lc_1=I$, $\Lc_2$ has to be equal to $\Lc_3$, and $Z_{I\,\Lc_2}^{\Lc_2}$ indicates a TDL in the space direction. When $\Lc_2=I$, there is a non-trivial TDL in the time direction.

The modular invariant partition function without the insertion of any TDLs is a sum of squares of sums of characters
\begin{equation}
Z_{II}^I=\sum_{a\;\in\;T}\chichi{a}{a},
\end{equation}
where $\chi_{b}=\sum_{i}n_{i1}^{b} \chi_{i}$ are the block characters and $T$ is a subset of all block characters.
Moreover, the subset $T$ is a subalgebra of the defect line fusion algebra $N$ \cite{Zuber_2000}.
From the property of TDLs given by Eq. \eqref{eq:Lchi}, one can show that
\begin{equation}\label{eq:Zb1b-def}
    Z_{bI}^b=\sum_{a\in T,c}N_{ab}^c\chichi{c}{a}.
\end{equation}
However, the fusion rule of topological defect lines $N_{ab}^c$ is unknown so far.
To proceed, we act with a modular $S$ transformation on $Z_{bI}^b$ to get $Z_{Ib}^b$
\begin{equation}\label{eq:Zb1b}\begin{aligned}
Z_{Ib}^b&=S(Z_{bI}^b)
=\sum_{a\in T,c}N_{ab}^c S(\chi_c)S_(\chibar_a)
=\sum_{a\in T;c}\sum_{i,j,k,l}N_{ab}^c n_{i1}^c n_{k1}^a S_{ij}S_{kl}^*\chichi{j}{l}.
\end{aligned}\end{equation}
If $N_{ab}^c$ is totally symmetric in all three indices and $T$ is the set of all block characters, $Z_{1b}^b$ can be simplified further
\begin{equation}\label{eq:Z1bb}\begin{aligned}
    Z_{Ib}^b&=\sum_{a,c}\sum_{i,j,k,l}N_{ac}^b n_{i1}^c n_{k1}^a S_{ij}S_{kl}^*\chichi{j}{l}
    =\sum_{i,j,k,l,m} N_{ik}^m n_{m1}^b S_{ij}S_{kl}^*\chichi{j}{l}\\
    &=\sum_{j,m} \frac{S_{mj}}{S^*_{1j}} n_{m1}^b \chichi{j}{j}~,
\end{aligned}\end{equation}
where the last equality comes from the Verlinda formula Eq. \eqref{eq:Verlinde-formula}.
The components such as $S$ and $n_{m1}^b$ in the last line are known for a given theory.
So we can use them to compute $Z_{Ib}^b$, then perform a modular $S$ transformation to get $Z_{bI}^b$ and extract $N_{ac}^b$ from it using Eq. \eqref{eq:Zb1b-def}.

If $N_{ab}^c$ is not totally symmetric in all three indices but $T$ is the set of all block characters, we notice that
\begin{equation}
    n_{i1}^{c}=n_{i1}^{\bar{c}},
\end{equation}
because $c$ and $\bar{c}$ are conjugate to each other, and so share the same character formula.
The derivation in Eq. \eqref{eq:Z1bb} works as well
\begin{equation}\label{eq:Z1bb-2}\begin{aligned}
Z_{Ib}^b&=\sum_{a,c}\sum_{i,j,k,l}N_{a\bar{c}}^{\bar{b}} n_{i1}^c n_{k1}^a S_{ij}S_{kl}^*\chichi{j}{l}
=\sum_{a,c}\sum_{i,j,k,l}N_{a\bar{c}}^{\bar{b}} n_{i1}^{\bar{c}} n_{k1}^a S_{ij}S_{kl}^*\chichi{j}{l}\\
&=\sum_{j,m} \frac{S_{mj}}{S^*_{1j}} n_{m1}^{\bar{b}} \chichi{j}{j}
=\sum_{j,m} \frac{S_{mj}}{S^*_{1j}} n_{m1}^b \chichi{j}{j}.
\end{aligned}\end{equation}
If $T$ is not the set of all block characters, let $T^\perp$ be the complement of $T$, we need to subtract from the above equation the term
\begin{equation}
    \sum_{a\in T^\perp}N_{ab}^c n_{i1}^c n_{j1}^a S_{ik}\chi_{k} S^*_{jl}\chibar_{l}
    \xrightarrow{S\textrm{ transform}}
    \sum_{a\in T^\perp}N_{ab}^c n_{i1}^c n_{j1}^a \chi_{i} \chibar_{j}
    =\sum_{a\in T^\perp}N_{ab}^c \chi_{c} \chibar_{a},
\end{equation}
to get the correct result of $Z_{Ib}^b$.

In the process of computing TDLs in the Parafermion theory, we propose to split some characters into ``sub-characters'' and add more characters to make the $S$ transformation close.
In this way, we can get a larger, extended $S$ matrix $\hat{S}$.
Then, we can use the above method to compute TDLs and their fusion rules
\begin{equation}\label{eq:generalVerlinde}
    \sum_{a\in T}\Nhat_{ab}^c\chichihat{c}{a}+\sum_{a\in T^\perp}\Nhat_{ab}^c \chichihat{c}{a}=\hat{S}\left(\sum_{j,m}\frac{\Shat_{mj}}{\Shat^*_{1j}}n_{m1}^b\chichihat{j}{j}\right)
\end{equation}
where the hat $\hat{\phantom{\chi}}$ denotes the extended version.
This is a generalization of the Verlinde formula.
For Parafermion theory, the characters that are required to split are the ones that appear in the sum of $\ZZ_k$ simple objects.
We will show this process using $\ZZ_3$ to $\ZZ_5$ Parafermion theory.


\subsection{The Ising Model}
An illustrative example is the Ising model.
On the other hand, the Ising model is also the $k=2$ Parafermion theory.
In the following, we will implement the above procedure to extract the topological defect lines of the Ising model.
The $S$ matrix for the Ising model is \cite{Moore:1988qv}
\begin{equation}
    S= \begin{psmallmatrix}
     \frac{1}{2} & \frac{1}{2} & \frac{1}{\sqrt{2}} \\
     \frac{1}{2} & \frac{1}{2} & \frac{-1}{\sqrt{2}} \\
     \frac{1}{\sqrt{2}} & \frac{-1}{\sqrt{2}} & 0 \\
    \end{psmallmatrix},
\end{equation}
in which the order of characters is $\vec{\chi}=(\chi_1,\chi_\epsilon,\chi_\sigma)^t$.
The conformal weights for $1,\epsilon$ and $\sigma$ are $0,\frac{1}{2}$ and $\frac{1}{16}$.
The fusion rules for the Ising primaries are
\begin{equation}
    \sigma\otimes\sigma=1\oplus\epsilon,\quad \sigma\otimes\epsilon=\sigma,\quad\epsilon\otimes\epsilon=1.
\end{equation}
Starting from its diagonal partition function
$Z_{11}^{1}=\vec{\chi}^t
\begin{psmallmatrix}
1\\&1\\&&1
\end{psmallmatrix}\vec{\chi}$,
we know that
\begin{equation}
    n_{i1}^{a}=\begin{psmallmatrix}
    1\\&1\\&&1\\
    \end{psmallmatrix}.
\end{equation}
Then by using Eq. \eqref{eq:Z1bb}, we get the expression for $Z_{1b}^b$, $b\in\{1, 2, 3\}$,
\begin{equation}
    Z_{11}^{1}=\begin{psmallmatrix}
         1 &   &   \\& 1 &   \\&   & 1 \\
    \end{psmallmatrix},\quad
    Z_{12}^{2}=\begin{psmallmatrix}
         1 &   &   \\& 1 &   \\&   & -1 \\
    \end{psmallmatrix},\quad
    Z_{13}^{3}=\sqrt{2}\begin{psmallmatrix}
         1 &   &   \\& -1 &   \\&   & 0 \\
    \end{psmallmatrix}.
\end{equation}
The hat on $\hat{1}, \hat{2}, \hat{3}$ emphasizes that they are indices for TDLs.
Acting then with $S$ on the above matrices, we get
\begin{equation}
    Z_{11}^{1}=\begin{psmallmatrix}
         1 &   &   \\& 1 &   \\&   & 1 \\
    \end{psmallmatrix},\quad
    Z_{21}^{2}=\begin{psmallmatrix}
           & 1 &   \\1 &   &   \\&   & 1 \\
    \end{psmallmatrix},\quad
    Z_{31}^{3}=\begin{psmallmatrix}
           &   & 1 \\ &   & 1 \\ 1 & 1 &   \\
    \end{psmallmatrix}.
\end{equation}
By comparing the expressions for $Z_{b1}^b$ with Eq. \eqref{eq:Zb1b-def}, we can read off the fusion rules of the defect lines
\begin{equation}
    3\otimes3=1\oplus2, \quad
    3\otimes2=3, \quad
    2\otimes2=1,
\end{equation}
which happen to be the same as the fusion rules of the primaries under the map $1\to1,\epsilon\to 2,\sigma\to 3$.
Now that we know both $n_{i1}^a$ and the fusion rules for defect lines $N_{ab}^c$, we can use $n_{ib}^c=n_{i1}^a N_{ba}^c$ to get all $n_{ia}^b$
\begin{equation}
    n_{i 1}^{a}=\begin{psmallmatrix}
    1\\&1\\&&1\\
    \end{psmallmatrix},\quad
    n_{i 2}^{a}=\begin{psmallmatrix}
    &1\\1\\&&1
    \end{psmallmatrix},\quad
    n_{i 3}^{a}=\begin{psmallmatrix}
    &&1\\&&1\\1&1
    \end{psmallmatrix}.
\end{equation}
Reindexing the above $n_{ia}^b$ gives
\begin{equation}
    n_{1 a}^{b}=\begin{psmallmatrix}
    1\\&1\\&&1\\
    \end{psmallmatrix},\quad
    n_{\epsilon a}^{b}=\begin{psmallmatrix}
    &1\\1\\&&1
    \end{psmallmatrix},\quad
    n_{\sigma a}^{b}=\begin{psmallmatrix}
    &&1\\&&1\\1&1
    \end{psmallmatrix}.
\end{equation}
One can explicitly check that the $n_i$ obey Eq. \ref{eq:nn=Nn},
\begin{equation}
    n_\epsilon n_\epsilon=n_1,\quad
    n_\sigma n_\epsilon = n_\epsilon n_\sigma = n_\sigma,\quad
    n_\sigma n_\sigma = n_1 + n_\epsilon.
\end{equation}

\subsection{Three-State Potts Model}\label{sec:potts}

The second example is the $c=\frac{4}{5}$ 3-state Potts Model.
This is the simplest case in which the number of defect lines does not equal the number of primaries.
We will first use the method in \cite{Chang:2018iay} to compute TDLs and then go back to the method in Section \ref{sec:basicsetups}.
We start from the non-diagonal partition function of the 3-state Potts model,
\small
\begin{equation}
    Z_{II}^{I}
    =\Vec{\chi}^\dagger\begin{psmallmatrix}
         1 & 1 &   &   &   &   &   &   &   &   \\
         1 & 1 &   &   &   &   &   &   &   &   \\
           &   & 1 & 1 &   &   &   &   &   &   \\
           &   & 1 & 1 &   &   &   &   &   &   \\
           &   &   &   & 2 &   &   &   &   &   \\
           &   &   &   &   & 2 &   &   &   &   \\
           &   &   &   &   &   & 0 &   &   &   \\
           &   &   &   &   &   &   & 0 &   &   \\
           &   &   &   &   &   &   &   & 0 &   \\
           &   &   &   &   &   &   &   &   & 0 \\
    \end{psmallmatrix}\Vec{\chi},
\end{equation}
\normalsize
where the primaries are ordered as 
\begin{equation}\label{eq:Potts-char-order}
\begin{array}{rcccccccccc}
\Vec{\chi}^t = &
\left(\chi_{1,1},\right.&\chi_{1,4},&\chi_{1,2},&\chi_{1,3},&\chi_{3,4},&\chi_{3,3},&\chi_{2,4},&\chi_{4,4},&\chi_{2,2},&\left.\chi_{2,3}\right)\\
=&\left(\chi_{I},\right.&\chi_{Y},&\chi_{\epsilon},&\chi_{X},&\chi_{Z},&\chi_{\sigma},&\chi_{2,4},&\chi_{4,4},&\chi_{2,2},&\left.\chi_{2,3}\right)~,
\end{array}
\end{equation}
in which the first line uses minimal model notation to denote characters $\chi_{r,s}$ and the second line uses traditional symbols.
The nonzero entries correspond to the 12 Virasoro primaries in this model, as listed below
\begin{equation}
    1_{0,0},\; \epsilon_{\frac{2}{5},\frac{2}{5}},\; X_{\frac{7}{5},\; \frac{7}{5}},Y_{3,3},\;
    \Omega_{3,0},\; \Tilde\Omega_{0,3},\; \Phi_{\frac{7}{5},\frac{2}{5}},\; \Tilde{\Phi}_{\frac{2}{5},\frac{7}{5}},\;
    Z_{\frac{2}{3},\frac{2}{3}},\; Z^*_{\frac{2}{3},\frac{2}{3}},\; \sigma_{\frac{1}{15},\frac{1}{15}},\; \sigma^*_{\frac{1}{15},\frac{1}{15}}.
\end{equation}
The $n_{i1}^a$ matrix in this theory is
\footnotesize
\begin{equation}\label{eq:Potts-n}
n_{i1}^a=
    \left(
\begin{array}{cccccccc}
 1 &   &   &   &   &   &   &   \\
 1 &   &   &   &   &   &   &   \\
   &   &   & 1 &   &   &   &   \\
   &   &   & 1 &   &   &   &   \\
   & 1 & 1 &   &   &   &   &   \\
   &   &   &   & 1 & 1 &   &   \\
   &   &   &   &   &   & 1 &   \\
   &   &   &   &   &   & 1 &   \\
   &   &   &   &   &   &   & 1 \\
   &   &   &   &   &   &   & 1 \\
\end{array}
\right),
\end{equation}
\normalsize
in which the rows stand for primaries ordered as Eq. \eqref{eq:Potts-char-order} and the columns stand for TDLs.
We name these 8 TDLs as $(I , \eta  , \eta ^2 , W , W\eta , W\eta^2 , N , WN)$.
Using the method developed in Eq. \eqref{eq:Z1bb-2}, we can compute the partition functions of these TDLs
\begin{equation}\label{eq:PottsZ}\begin{aligned}
& Z_{I}=\begin{psmallmatrix}
 1 &   &   &   &   &   \\
   & 1 &   &   &   &   \\
   &   & 2 &   &   &   \\
   &   &   & 2 &   &   \\
   &   &   &   & 0 &   \\
   &   &   &   &   & 0 \\
\end{psmallmatrix},
Z_{\eta}=Z_{\eta\eta}=\begin{psmallmatrix}
   &   & 1 &   &   &   \\
   &   &   & 1 &   &   \\
 1 &   & 1 &   &   &   \\
   & 1 &   & 1 &   &   \\
   &   &   &   & 0 &   \\
   &   &   &   &   & 0 \\
\end{psmallmatrix},\\
&Z_{W}=\begin{psmallmatrix}
   & 1 &   &   &   &   \\
 1 & 1 &   &   &   &   \\
   &   &   & 2 &   &   \\
   &   & 2 & 2 &   &   \\
   &   &   &   & 0 &   \\
   &   &   &   &   & 0 \\
\end{psmallmatrix},
Z_{W\eta}=Z_{W\eta \eta}=\begin{psmallmatrix}
   &   &   & 1 &   &   \\
   &   & 1 & 1 &   &   \\
   & 1 &   & 1 &   &   \\
 1 & 1 & 1 & 1 &   &   \\
   &   &   &   & 0 &   \\
   &   &   &   &   & 0 \\
\end{psmallmatrix},\\
&Z_N=\begin{psmallmatrix}
 0 &   &   &   & 1 &   \\
   & 0 &   &   &   & 1 \\
   &   & 0 &   & 2 &   \\
   &   &   & 0 &   & 2 \\
   &   &   &   & 0 &   \\
   &   &   &   &   & 0 \\
\end{psmallmatrix},
Z_{WN}=\begin{psmallmatrix}
 0 &   &   &   &   & 1 \\
   & 0 &   &   & 1 & 1 \\
   &   & 0 &   &   & 2 \\
   &   &   & 0 & 2 & 2 \\
   &   &   &   & 0 &   \\
   &   &   &   &   & 0 \\
\end{psmallmatrix},
\end{aligned}\end{equation}
where we condense the basis as $(\chi_{I}+\chi_{Y}, \chi_{\epsilon}+\chi_{X}, \chi_{Z}, \chi_{\sigma}, \chi_{2,4}+\chi_{4,4}, \chi_{2,2}+\chi_{2,3})$ to make the matrix smaller.
These are the partition functions that were given at the end of \cite{Petkova:2000ip}.
The fusion rule is computed as follows.
To compute the fusion result of $a$ and $b$, one can first compute the space direction partition function $Z_{I\,ab}^{ab}$, then decompose it in the basis of all TDLs $Z_{I\,a}^{a}$.
The decomposition factor is the fusion rule.
The fusion rule of 3-state Potts model TDLs is generated by
\begin{equation}\begin{aligned}
    \eta\otimes\eta=\eta^2,\;\eta\otimes\eta^2=I,\;
    W\otimes W=I\oplus W,\\
    N\otimes N=I\oplus\eta\oplus\eta^2,\;
    \eta\otimes N=N.
\end{aligned}
\end{equation}
This fusion algebra is the minimal fusion algebra that is larger than the fusion algebra of Verlinde lines.
More explicitly, the fusion rule is
\small
\begin{equation}\label{eq:Potts-fusionrule}
\begin{array}{c|ccccccc}
I & \eta  & \eta ^2 & W & W\eta & W\eta^2 & N & WN \\
\hline
\eta  & \eta ^2 & I & W\eta & W\eta^2 & W & N & WN \\
\eta ^2 & I & \eta  & W\eta^2 & W & W\eta & N & WN \\
W & W\eta & W\eta^2 & I\oplus W & \eta\oplus W\eta & \eta^2\oplus W\eta^2 & WN & N\oplus WN \\
W\eta & W\eta^2 & W & \eta\oplus W\eta & \eta^2\oplus W\eta^2 & I\oplus W & WN & N\oplus WN \\
W\eta^2 & W & W\eta & \eta^2+W\eta^2 & I\oplus W & \eta\oplus W\eta & WN & N\oplus WN \\
N & N & N & WN & WN & WN & I\oplus\eta\oplus\eta^2 & W\oplus W\eta\oplus W\eta^2 \\
WN & WN & WN & N\oplus WN & N\oplus WN & N\oplus WN & W\oplus W\eta\oplus W\eta^2 & \begin{matrix}
    I\oplus\eta\oplus\eta^2\oplus\\
    W\oplus W\eta\oplus W\eta^2
\end{matrix}\\
\end{array}
\end{equation}
\normalsize
One can explicitly check that the above fusion rule satisfies Eq. \eqref{eq:NN=NN}, 
and so a valid fusion algebra.
The fusion rule $N$ is not totally symmetric. As $\sigma$ is dual to $\sigma^*$ and $Z$ to $Z^*$, 
\begin{equation}
    N_{\sigma}=N_{\sigma^*}^t,\quad N_{Z}=N_{Z^*}^t.
\end{equation}
This is the property required by Eq. \eqref{eq:Nsym}.

\subsection[Z3 Parafermion Theory]{$\ZZ_3$ Parafermion Theory}

The $\ZZ_3$ Parafermion theory shares the same central charge $c=\frac{4}{5}$ with the 3-state Potts model.
It has 6 primaries denoted by $\psi_{3,3},\psi_{3,-1},\psi_{3,1},\psi_{2,0},\psi_{1,1},\psi_{2,2}$.
The conformal weights of the primaries are shown in Table \ref{tab:wt-z3}.
To compute the TDLs, we find that we have to split the characters $\chi_{3,3}$ and $\chi_{2,0}$ into two ``sub-characters''.
Then we can use the method in Section \ref{sec:basicsetups}, especially Eq. \eqref{eq:Zb1b} and \eqref{eq:Z1bb-2} to compute the TDLs.
During the S transformation, extra characters get involved to make the algebra close.
As a result, the S transformation represented as a matrix is larger than the original S matrix.
The splitting happens to be the same as Eq. \eqref{eq:Z3-Potts-char}.
So we can make use of the given result in the 3-state Potts model.
Actually, the new $S$ matrix is exactly the same as that of the 3-state Potts model
More explicitly, the extended $S$ matrix is
\begin{equation}
    \hat{S}=\sqrt{\frac{1}{15}}\begin{pmatrix}
s_1& s_1 & s_2 & s_2 & 2s_1 & 2s_2 & \sqrt{3}s_1 & \sqrt{3}s_1 & \sqrt{3}s_2 & \sqrt{3}s_2 \\
s_1& s_1 & s_2 & s_2 & 2s_1 & 2s_2 & -\sqrt{3}s_1 & -\sqrt{3}s_1 & -\sqrt{3}s_2 & -\sqrt{3}s_2 \\
s_2& s_2 & -s_1 & -s_1 & 2s_2 & -2s_1 & -\sqrt{3}s_2 & -\sqrt{3}s_2 & \sqrt{3}s_1 & \sqrt{3}s_1 \\
s_2& s_2 & -s_1 & -s_1 & 2s_2 & -2s_1 & \sqrt{3}s_2 & \sqrt{3}s_2 & -\sqrt{3}s_1 & -\sqrt{3}s_1 \\
2s_1 & 2s_1 & 2s_2 & 2s_2 & -2s_1 & -2s_2 & 0 & 0 & 0 & 0 \\
2s_2 & 2s_2 & -2s_1 & -2s_1 & -2s_2 & 2s_1 & 0 & 0 & 0 & 0 \\
\sqrt{3}s_1 & -\sqrt{3}s_1 & -\sqrt{3}s_2 & \sqrt{3}s_2 & 0 & 0 & -\sqrt{3}s_1 & \sqrt{3}s_1 & -\sqrt{3}s_2 & \sqrt{3}s_2 \\
\sqrt{3}s_1 & -\sqrt{3}s_1 & -\sqrt{3}s_2 & \sqrt{3}s_2 & 0 & 0 & \sqrt{3}s_1 & -\sqrt{3}s_1 & \sqrt{3}s_2 & -\sqrt{3}s_2 \\
\sqrt{3}s_2 & -\sqrt{3}s_2 & \sqrt{3}s_1 & -\sqrt{3}s_1 & 0 & 0 & -\sqrt{3}s_2 & \sqrt{3}s_2 & \sqrt{3}s_1 & -\sqrt{3}s_1 \\
\sqrt{3}s_2 & -\sqrt{3}s_2 & \sqrt{3}s_1 & -\sqrt{3}s_1 & 0 & 0 & \sqrt{3}s_2 & -\sqrt{3}s_2 & -\sqrt{3}s_1 & \sqrt{3}s_1
    \end{pmatrix},
\end{equation}
where $s_1=\sin{\frac{\pi}{5}}$ and $s_2=\sin{\frac{2\pi}{5}}$.
The characters are ordered as $\Vec{\chi}=(\chi_{3,3}^1,\chi_{3,3}^2,\chi_{2,0}^1,\chi_{2,0}^2,\chi_{3,1}=\chi_{3,-1},\chi_{1,1}=\chi_{2,2},\textrm{4 more characters to make S close})$.
The last two groups of characters always appear together.
So we can treat the sum of them as one character to write down a smaller $S$ matrix
\begin{equation}
    \hat{S}'=\sqrt{\frac{1}{15}}\begin{pmatrix}
s_1& s_1 & s_2 & s_2 & 2s_1 & 2s_2 & \sqrt{3}s_1 & \sqrt{3}s_2  \\
s_1& s_1 & s_2 & s_2 & 2s_1 & 2s_2 & -\sqrt{3}s_1 & -\sqrt{3}s_2 \\
s_2& s_2 & -s_1 & -s_1 & 2s_2 & -2s_1 & -\sqrt{3}s_2 & \sqrt{3}s_1 \\
s_2& s_2 & -s_1 & -s_1 & 2s_2 & -2s_1 & \sqrt{3}s_2 & -\sqrt{3}s_1 \\
2s_1 & 2s_1 & 2s_2 & 2s_2 & -2s_1 & -2s_2 & 0 & 0 \\
2s_2 & 2s_2 & -2s_1 & -2s_1 & -2s_2 & 2s_1 & 0 & 0 \\
2\sqrt{3}s_1 & -2\sqrt{3}s_1 & -2\sqrt{3}s_2 & 2\sqrt{3}s_2 & 0 & 0 & 0 & 0 \\
2\sqrt{3}s_2 & -2\sqrt{3}s_2 & 2\sqrt{3}s_1 & -2\sqrt{3}s_1 & 0 & 0 & 0 & 0
    \end{pmatrix}.
\end{equation}

The solution of TDLs is the same as the one in the previous section.
Among all the TDLs, the most interesting are the ones that form a TY category subset.
In the case of the $\ZZ_3$ Parafermions, $I,\eta,\eta^2$ clearly form a Abelian $\ZZ_3$ fusion algebra and $N$ is the non-invertible object
\begin{equation}
    N\otimes N=I\oplus\eta\oplus\eta^2.
\end{equation}

A hint of why $\chi_{3,3}$ and $\chi_{2,0}$ are required to be split lies in the partition function of $N\otimes N$. There exists a TY category in the Parafermion theory and the non-invertible object $N$ fused to itself equals the sum of all $\ZZ_k$ simple objects, in this case, the Verlinde lines corresponding to $\psi_{3,3},\psi_{3,1},\psi_{3,1}$ or $1,\eta,\eta^2$. In fact,
\begin{equation}
Z_{II}^I+Z_{I\eta}^\eta+Z_{I\eta^2}^{\eta^2}=3\chi_{3,3}\chibar_{3,3}+3\chi_{2,0}\chibar_{2,0}.
\end{equation}
This hints towards splitting $\chi_{3,3}$ and $\chi_{2,0}$ to get the correct action for the $N$ defect line.

\subsection[Z4 Parafermion Theory]{$\ZZ_4$ Parafermion Theory}

The $\ZZ_4$ Parafermion theory has $c=1$ and 10 primaries 
$\psi_{4,4}$, $\psi_{4,0}$, $\psi_{4,2}$, $\psi_{4,-2}$, $\psi_{2,0}$, $\psi_{2,2}$, $\psi_{3,-1}$, $\psi_{3,1}$, $\psi_{3,3}$ and $\psi_{1,1}$ where the subscripts of $\psi_{l,m}$, given by $l$ and $m$, label the spin of the field.
Their conformal weights are shown in Table \ref{tab:wt-z4}.
\begin{table}[h]
    \centering
    \begin{tabular}{c|ccccccc}
        $m$&-2&-1&0&1&2&3&4\\
        \hline
        $l=1$&&&& $\frac{1}{16}$\\
        $l=2$&&& $\frac{1}{3}$&& $\frac{1}{12}$\\
        $l=3$&& $\frac{9}{16}$&& $\frac{9}{16}$&& $\frac{1}{16}$\\
        $l=4$& $\frac{3}{4}$&& 1&& $\frac{3}{4}$ && $0$
    \end{tabular}
    \caption{Conformal weights for $\ZZ_4$ Parafermion primaries. There are 10 primaries.}
    \label{tab:wt-z4}
\end{table}

As the primaries of the same conformal weight share the same character formula, there are 7 distinct holomorphic or antiholomorphic characters in the $\ZZ_4$ Parafermion theory.
They are
\begin{equation}\begin{aligned}
\chi_{4,4}&=\frac{1}{2\eta}\sum_{k}(-1)^k q^{k^2} +q^{3k^2},\qquad
\chi_{4,0}=\frac{1}{2\eta}\sum_{k}-(-1)^k q^{k^2} +q^{3k^2},\\
\chi_{4,2}&=\chi_{4,-2}=\frac{1}{2\eta}\sum_{k=3,9\textrm{ mod }12}q^{k^2/12},\\
\chi_{2,0}&=\frac{1}{2\eta}\sum_{k}2q^{(6k+2)^2/12}=\frac{1}{2\eta}\sum_{k=2,4,8,10 \textrm{ mod } 12}q^{k^2/12},\\
\chi_{2,2}&=\frac{1}{2\eta}\sum_{k}2q^{(6k+1)^2/12}=\frac{1}{2\eta}\sum_{k=1,5,7,11 \textrm{ mod }12}q^{k^2/12},\\
\chi_{3,3}&=\chi_{1,1}=\frac{1}{2\eta}\sum_{k=1,7,9,15 \textrm{ mod }16}q^{k^2/16},\qquad
\chi_{3,-1}=\chi_{3,1}=\frac{1}{2\eta}\sum_{k=3,5,11,13 \textrm{ mod }16}q^{k^2/16},
\end{aligned}
\end{equation}
where $q\equiv e^{2\pi i \tau}$ and $\eta$ is the Dedekind eta function.
The $S$ matrix can be computed by Eq. \eqref{eq:pafaS}.
To compute the TDLs, proceeding analogously to the $\ZZ_3$ case, we need to split some of the characters.
As suggested by \cite{Thorngren:2021yso}, by the argument of $\ZZ_4$ Parafermion theory being dual to the $\ZZ_2^C$ orbifold of compact boson theory $U(1)_{6}$, denoted as $U(1)_6/\ZZ^C_2$, 
the characters need to be split into a two-dimensional state space based on the spin 3 generator under the duality.
It turns out that $\chi_{4,4}$, $\chi_{4,0}$ and $\chi_{2,0}$ are split into two sub-characters as
\begin{equation}\begin{aligned}
\chi_{4,4}&=\frac{1}{2\eta}\left(\sum_{k}(-1)^k q^{k^2} +\sum_{k\textrm{ even}}q^{3k^2}\right)+\frac{1}{2\eta}\left(\sum_{k\textrm{ odd}}q^{3k^2}\right)
:=\chi_{4,4}'+\chi_{4,4}''\\
\chi_{4,0}&=\frac{1}{2\eta}\left(-\sum_{k}(-1)^k q^{k^2} +\sum_{k\textrm{ even}}q^{3k^2}\right)+\frac{1}{2\eta}\left(\sum_{k\textrm{ odd}}q^{3k^2}\right)
:=\chi_{4,0}'+\chi_{4,0}''\\
\chi_{2,0}&=\frac{1}{2\eta}\left(\sum_{k\textrm{ even}}2q^{(6k+2)^2/12}\right)+\frac{1}{2\eta}\left(\sum_{k\textrm{ odd}}2q^{(6k+2)^2/12}\right)
:=\chi_{2,0}'+\chi_{2,0}''.
\end{aligned}
\end{equation}
That $\chi_{4,4}$, $\chi_{4,0}$ and $\chi_{2,0}$ need to be split can be also seen from the expression of the sum of all $\ZZ_k$ simple objects
\begin{equation}
Z_{II}^{I}+Z_{I\eta}^{\eta}+Z_{I\eta^2}^{\eta^2}+Z_{I\eta^3}^{\eta^3}=4\chichi{4,4}{4,4}+4\chichi{4,0}{4,0}+4\chichi{2,0}{2,0}.
\end{equation}
To make the $S$ transform close, the extended $S$ matrix contains three more rows and columns.
The three new characters are
\begin{equation}\begin{aligned}
\Shat(\chi_{4,4}'')&=\Shat(\chi_{4,0}'')=\frac{1}{4\sqrt{6}}\left(\chi_{4,4}+\chi_{4,2}+\chi_{4,0}+\chi_{4,-2}+2\chi_{2,0}+2\chi_{2,2}-2\frac{1}{2\eta}\sum_{k \textrm{ odd}}q^{k^2/48}\right),\\
\Shat(\chi_{2,0}'')&=\frac{1}{2\sqrt{6}}\left(\chi_{4,4}+\chi_{4,2}+\chi_{4,0}+\chi_{4,-2}-\chi_{2,0}-\chi_{2,2}\right)\\
&-\frac{1}{2\sqrt{6}}\frac{1}{2\eta}\sum_{k}\left(q^{(6k+1)^2/48}-q^{(6k+3)^2/48}\right).
\end{aligned}\end{equation}
The fusion rule for $\ZZ_4$ Parafermion Verlinde lines is generated by $\psi_{4,4}\equiv I$, $\psi_{4,2}\equiv\eta$, $\psi_{2,0}\equiv V$ and $\psi_{3,1}\equiv W$ obeying
\begin{equation}
    \eta^4 = I,\quad V\otimes V=I\oplus\eta^2\oplus V,\quad W\otimes V=W\oplus W\eta^2,\quad W\otimes W=\eta^3\oplus V\eta.
\end{equation}
The fusion table is
\footnotesize
\begin{equation}\label{eq:Z4-fusion}
\begin{array}{c|ccccccccc}
\text{I} & \eta  & \eta ^2 & \eta ^3 & \text{V} & \text{V$\eta $} & \text{W} & \text{W$\eta $} & \text{W$\eta $}^2 & \text{W$\eta $}^3 \\
 \hline
 \eta  & \eta ^2 & \eta ^3 & \text{I} & \text{V$\eta $} & \text{V} & \text{W$\eta $} & \text{W$\eta $}^2 & \text{W$\eta $}^3 & \text{W} \\
 \eta ^2 & \eta ^3 & \text{I} & \eta  & \text{V} & \text{V$\eta $} & \text{W$\eta $}^2 & \text{W$\eta $}^3 & \text{W} & \text{W$\eta $} \\
 \eta ^3 & \text{I} & \eta  & \eta ^2 & \text{V$\eta $} & \text{V} & \text{W$\eta $}^3 & \text{W} & \text{W$\eta $} & \text{W$\eta $}^2 \\
 \text{V} & \text{V$\eta $} & \text{V} & \text{V$\eta $} & \text{I}\oplus\eta ^2\oplus\text{V} & \eta ^3\oplus\eta \oplus\text{V$\eta $} & \text{W}\oplus\text{W$\eta $}^2 & \text{W$\eta $}^3\oplus\text{W$\eta $} & \text{W}\oplus\text{W$\eta $}^2 & \text{W$\eta $}^3\oplus\text{W$\eta $} \\
 \text{V$\eta $} & \text{V} & \text{V$\eta $} & \text{V} & \eta ^3\oplus\eta \oplus\text{V$\eta $} & \text{I}\oplus\eta ^2\oplus\text{V} & \text{W$\eta $}^3\oplus\text{W$\eta $} & \text{W}\oplus\text{W$\eta $}^2 & \text{W$\eta $}^3\oplus\text{W$\eta $} & \text{W}\oplus\text{W$\eta $}^2 \\
 \text{W} & \text{W$\eta $} & \text{W$\eta $}^2 & \text{W$\eta $}^3 & \text{W}\oplus\text{W$\eta $}^2 & \text{W$\eta $}^3\oplus\text{W$\eta $} & \eta ^3\oplus\text{V$\eta $} & \text{I}\oplus\text{V} & \eta \oplus\text{V$\eta $} & \eta ^2\oplus\text{V} \\
 \text{W$\eta $} & \text{W$\eta $}^2 & \text{W$\eta $}^3 & \text{W} & \text{W$\eta $}^3\oplus\text{W$\eta $} & \text{W}\oplus\text{W$\eta $}^2 & \text{I}\oplus\text{V} & \eta \oplus\text{V$\eta $} & \eta ^2\oplus\text{V} & \eta ^3\oplus\text{V$\eta $} \\
 \text{W$\eta $}^2 & \text{W$\eta $}^3 & \text{W} & \text{W$\eta $} & \text{W}\oplus\text{W$\eta $}^2 & \text{W$\eta $}^3\oplus\text{W$\eta $} & \eta \oplus\text{V$\eta $} & \eta ^2\oplus\text{V} & \eta ^3\oplus\text{V$\eta $} & \text{I}\oplus\text{V} \\
 \text{W$\eta $}^3 & \text{W} & \text{W$\eta $} & \text{W$\eta $}^2 & \text{W$\eta $}^3\oplus\text{W$\eta $} & \text{W}\oplus\text{W$\eta $}^2 & \eta ^2\oplus\text{V} & \eta ^3\oplus\text{V$\eta $} & \text{I}\oplus\text{V} & \eta \oplus\text{V$\eta $}
\end{array}
\end{equation}
\normalsize
in which the notation implies the following representation
\begin{equation}
\begin{array}{c|cccccccccc}
\textrm{Symbol} & I & \eta & \eta^2 & \eta^3 & V & V\eta & W & W\eta & W\eta^2 & W\eta^3 \\
\hline
\textrm{Verlinde line of } & \psi_{4,4} & \psi_{4,2} & \psi_{4,0} & \psi_{4,-2} & \psi_{2,0} & \psi_{2,2} & \psi_{3,1} & \psi_{3,-1} & \psi_{1,1} & \psi_{3,3}
\end{array}~.
\end{equation}

Of most interest is the $N$ object in the TY category.
The topological defect line corresponding to
\begin{equation}
    Z_{NI}^{N}=2\chi_{4,4}(\overline{\chi}_{4,4}'-\overline{\chi}_{4,4}'')
    +2\chi_{4,0}(\overline{\chi}_{4,0}'-\overline{\chi}_{4,0}'')-2\chi_{2,0}(\overline{\chi}_{2,0}'-\overline{\chi}_{2,0}''),
\end{equation}
plays the role of $N$ here.
One can use the method in section \ref{sec:potts} to compute the fusion rules and confirm that
\begin{equation}
    N\otimes \eta = N,\quad
    N\otimes N= I\oplus\eta\oplus\eta^2\oplus\eta^3.
\end{equation}

\subsection[Z5 Parafermion Theory]{$\ZZ_5$ Parafermion Theory}

\begin{table}[ht]
\centering
\begin{tabular}{c|ccccccccccccc}
$m$  &-3&-2&-1&0&1&2&3&4&5\\
\hline
$l=1$&&&&& $\frac{2}{35}$\\
$l=2$&&&& $\frac{2}{7}$&& $\frac{3}{35}$\\
$l=3$&&& $\frac{17}{35}$&& $\frac{17}{35}$&& $\frac{3}{35}$\\
$l=4$&&$\frac{23}{35}$&&$\frac{6}{7}$&&$\frac{23}{35}$&&$\frac{2}{35}$\\
$l=5$&$\frac{4}{5}$&&$\frac{6}{5}$&&$\frac{6}{5}$&&$\frac{4}{5}$&&$0$\\
\end{tabular}
\caption{Conformal weights for $\ZZ_5$ Parafermion primaries. There are 15 primaries in the $\ZZ_5$ Parafermion theory, among them there are 9 different conformal weights/characters.}
\label{tab:wt-z5}
\end{table}

The $\ZZ_5$ Parafermion theory has 15 primaries. Their $l,m$ indices together with corresponding conformal weights are shown in Table \ref{tab:wt-z5}.
The Verlinde lines corresponding to $\psi_{5,5}$, $\psi_{5,3}$, $\psi_{5,1}$, $\psi_{5,-1}$ and $\psi_{5,-3}$ form a $\ZZ_5$ fusion algebra, whose simple objects we denote by $I,\eta,\eta^2,\eta^3,\eta^4$.
The remaining Verlinde lines are generated by $\chi_{4,0}:=W$ and $\chi_{2,0}:=V$ with the fusion rule
\begin{equation}
W\otimes W=I\oplus V,\quad W\otimes V=W\oplus V,\quad V\otimes V=I\oplus W\oplus V.
\end{equation}
The remaining simple objects are
\begin{equation}
    \begin{array}{c|cccccccccc}
        \textrm{Symbol} & W & W\eta & W\eta^2 & W\eta^3 & W\eta^4 &
        V & V\eta & V\eta^2 & V\eta^3 & V\eta^4 \\
        \hline
        \textrm{Verlinde line of } & \psi_{4,0} & \psi_{4,-2} & \psi_{1,1} & \psi_{4,4} & \psi_{4,2} &
        \psi_{2,0} & \psi_{3,3} & \psi_{3,1} & \psi_{3,-1} & \psi_{2,2}
    \end{array}
\end{equation}
From the partition function of the sum of all $\ZZ_k$ simple objects
\begin{equation}\label{eq:Z5-split}
    \sum_{i=0}^{4}Z_{I\eta^i}^{\eta^i}=5\chichi{5,5}{5,5}+5\chichi{4,0}{4,0}+5\chichi{2,0}{2,0},
\end{equation}
we know that $\chi_{5,5},\chi_{4,0}$ and $\chi_{2,0}$ should be split into two sub-characters.
A  key point of our method is that the actual knowledge about how characters split into two new characters is not necessary.
The only input information is that three characters, $\chi_{5,5},\chi_{4,0}$ and $\chi_{2,0}$, are split into sub-characters, as well as the fact that one needs to add three more characters to make the $S$ transformation close.
There are still unfixed degrees of freedom in the extended $S$ matrix.
They come from the choice of basis for the 3 extra characters arising due to the $S$ transformation.
Contrary to the cases of $\ZZ_3$ and $\ZZ_4$ Parafermion theories, we do not have a dual theory to compare to. To fix the extended $S$ matrix, we apply the method in Section \ref{sec:basicsetups} and require the fusion rules to be non-negative integers.
We shall leave the details of bootstrap in Appendix \ref{sec:appendZ5} as the matrices are too large.
This way we obtain a fusion algebra with $15+3$ simple objects.
This is also the minimal modification of the original Verlinde fusion algebra which contains a TY category.
The three new simple objects are denoted by $N$, $WN$, and $VN$. Apart from the Verlinde fusion rules, their fusion rule is generated by
\begin{equation}
    N\otimes N=\bigoplus_{i=0}^{4}\eta^i =I\oplus\eta\oplus\eta^2\oplus\eta^3\oplus\eta^4,\quad
    V\otimes N=VN,\quad W\otimes N=WN.
\end{equation}
Clearly, this $N$ combined with $I$ to $\eta^4$ forms a TY category.
A representative subset of TDLs fuses as follows
\footnotesize	
\begin{equation}\label{eq:Z5-fusionrule}
\centering
\begin{array}{c|ccccccccc}
I & \eta & \eta^2 & \eta^3 & \eta^4 & W & V & N & WN & VN \\
\hline
\eta & \eta^2 & \eta^3 & \eta^4 & I & W\eta & V\eta & N & WN & VN \\
\eta^2 & \eta^3 & \eta^4 & I & \eta & W\eta^2 & V\eta^2 & N & WN & VN \\
\eta^3 & \eta^4 & I & \eta & \eta^2 & W\eta^3 & V\eta^3 & N & WN & VN \\
\eta^4 & I & \eta & \eta^2 & \eta^3 & W\eta^4 & V\eta^4 & N & WN & VN \\
W & W\eta & W\eta^2 & W\eta^3 & W\eta^4 & I\oplus V & W\oplus V & WN & N\oplus VN & WN\oplus VN \\
V & V\eta & V\eta^2 & V\eta^3 & V\eta^4 & W\oplus V & I\oplus W\oplus V & VN & WN\oplus VN & N\oplus WN\oplus VN\\
N & N & N & N & N & WN & VN &
\bigoplus\limits_{i=0}^{4}\eta^i &
W\bigoplus\limits_{i=0}^{4}\eta^i &
V\bigoplus\limits_{i=0}^{4}\eta^i \\
WN & WN & WN & WN & WN & N\oplus VN & WN\oplus VN &
W\bigoplus\limits_{i=0}^{4}\eta^i &
(I\oplus V)\bigoplus\limits_{i=0}^{4}\eta^i &
(W\oplus V)\bigoplus\limits_{i=0}^{4}\eta^i \\
VN & VN & VN & VN & VN & WN\oplus VN & N\oplus WN\oplus VN & 
V\bigoplus\limits_{i=0}^{4}\eta^i &
(W\oplus V) \bigoplus\limits_{i=0}^{4}\eta^i &
(I\oplus W\oplus V) \bigoplus\limits_{i=0}^{4}\eta^i
\end{array}
\end{equation}
\normalsize

\section{Conclusion}
\label{sec:conclusions}

In this paper, we have introduced a new method for computing topological defect lines in 2d RCFTs.
We demonstrated the feasibility of the method by applying it to bosonized $\mathbb{Z}_k$ Parafermionic CFTs for the cases $k=2$ to $k=5$. The method is based on previous approaches in the literature but relies crucially on finding an \textit{extended} $S$ matrix to incorporate the action of the duality defect on the Hilbert space. The corresponding splitting of the characters was known for $k=3$ due to the relation between $\mathbb{Z}_3$ Parafermions and the 3-state Potts model, and for $k=4$ due to the relation of $\mathbb{Z}_4$ Parafermions to the compact boson CFT $U(1)_6/\ZZ_2^C$. What is new is that this splitting can be generalized to the $k=5$ case and the TDLs together with their fusion rules can be extracted easily. More interestingly, the actual modular forms into which the characters split don't have to be computed for the method to work, only the extended S-matrix itself is relevant. 

There are, however, some subtle points to consider. For the case $k=4$, namely the theory arising from orbifolding the $c=1$ compact boson CFT at radius $R=\sqrt{6}$, the method does not give the full set of fusion rules. In particular, the TY subcategory can be extracted, but a close fusion rule is hidden. One reason this could be traced back to is that the $\mathbb{Z}_{2k}$ Parafermionic theories have a further $\mathbb{Z}_2$ abelian subcategory which can be gauged independently. As pointed out for example in Section \ref{sec:fermionization}, this $\mathbb{Z}_2$ symmetry can be also used to perform Fermionization, and the corresponding target theory will inherit a TY-subcategory. The $\mathbb{Z}_2$-symmetry means that there will be TDLs which only furnish a $\mathbb{Z}_2$-orbit (and not a $\mathbb{Z}_4$-orbit) under the action of the generator of the abelian $\mathbb{Z}_4$-symmetry as can be inferred from Eq. \eqref{eq:Z4-fusion}. This also implies that when constructing the extended $S$ matrix, one has to take care of these $\mathbb{Z}_2$-subcategories. The correct way to go about this is to decompose the $\mathbb{Z}_k$ Parafermionic characters, which are $\mathcal{W}_k$ characters, into Virasoro characters. We leave this task to the future. 

Yet from another perspective, it would be very interesting to construct the modular forms involved in the character splitting explicitly. This task is related to the task of finding an appropriate mathematical basis for the Parafermionic characters which are related to indefinite theta functions.


\acknowledgments

We would like to thank Chiming Chang, Jin Chen, Xia Gu, Qingrui Wang, and J.-B. Zuber for their valuable discussions. BH would also like to thank the Max-Planck Institute for Mathematics in Bonn and the Theoretical Physics Department at Utrecht University, where part of this work was completed, for their hospitality.
YR is grateful to the Physics Department of Purdue University for the hospitality where part of this work was completed.
This work is supported by the Young Thousand Talents grant of China as well as by the NSFC grant 12250610187.


\appendix
\section{$\Shat$ of $Z_5$ Parafermion Theory}
\label{sec:appendZ5}
The original $S$ matrix for $Z_5$ Parafermion theory is $\sqrt{35}S=$
\tiny
\begin{equation}
\begin{psmallmatrix}
 2 s_1 & 2 s_1 & 2 s_1 & 2 s_1 & 2 s_1 & 2 s_3 & 2 s_3 & 2 s_3 & 2 s_3 & 2 s_3 & 2 s_2 & 2 s_2 & 2 s_2 & 2 s_2 & 2 s_2 \\
 2 s_1 & 2 \omega^4 s_1 & -2 \omega^3 s_1 & 2 \omega^2 s_1 & -2 \omega s_1 & 2 s_3 & 2 \omega^4 s_3 & -2 \omega^3 s_3 & 2 \omega^2 s_3 & -2 \omega s_3 & 2 s_2 & 2 \omega^4 s_2 & -2 \omega^3 s_2 & 2 \omega^2 s_2 & -2 \omega s_2 \\
 2 s_1 & -2 \omega^3 s_1 & -2 \omega s_1 & 2 \omega^4 s_1 & 2 \omega^2 s_1 & 2 s_3 & -2 \omega^3 s_3 & -2 \omega s_3 & 2 \omega^4 s_3 & 2 \omega^2 s_3 & 2 s_2 & -2 \omega^3 s_2 & -2 \omega s_2 & 2 \omega^4 s_2 & 2 \omega^2 s_2 \\
 2 s_1 & 2 \omega^2 s_1 & 2 \omega^4 s_1 & -2 \omega s_1 & -2 \omega^3 s_1 & 2 s_3 & 2 \omega^2 s_3 & 2 \omega^4 s_3 & -2 \omega s_3 & -2 \omega^3 s_3 & 2 s_2 & 2 \omega^2 s_2 & 2 \omega^4 s_2 & -2 \omega s_2 & -2 \omega^3 s_2 \\
 2 s_1 & -2 \omega s_1 & 2 \omega^2 s_1 & -2 \omega^3 s_1 & 2 \omega^4 s_1 & 2 s_3 & -2 \omega s_3 & 2 \omega^2 s_3 & -2 \omega^3 s_3 & 2 \omega^4 s_3 & 2 s_2 & -2 \omega s_2 & 2 \omega^2 s_2 & -2 \omega^3 s_2 & 2 \omega^4 s_2 \\
 2 s_3 & 2 s_3 & 2 s_3 & 2 s_3 & 2 s_3 & -2 s_2 & -2 s_2 & -2 s_2 & -2 s_2 & -2 s_2 & 2 s_1 & 2 s_1 & 2 s_1 & 2 s_1 & 2 s_1 \\
 2 s_3 & 2 \omega^4 s_3 & -2 \omega^3 s_3 & 2 \omega^2 s_3 & -2 \omega s_3 & -2 s_2 & -2 \omega^4 s_2 & 2 \omega^3 s_2 & -2 \omega^2 s_2 & 2 \omega s_2 & 2 s_1 & 2 \omega^4 s_1 & -2 \omega^3 s_1 & 2 \omega^2 s_1 & -2 \omega s_1 \\
 2 s_3 & -2 \omega^3 s_3 & -2 \omega s_3 & 2 \omega^4 s_3 & 2 \omega^2 s_3 & -2 s_2 & 2 \omega^3 s_2 & 2 \omega s_2 & -2 \omega^4 s_2 & -2 \omega^2 s_2 & 2 s_1 & -2 \omega^3 s_1 & -2 \omega s_1 & 2 \omega^4 s_1 & 2 \omega^2 s_1 \\
 2 s_3 & 2 \omega^2 s_3 & 2 \omega^4 s_3 & -2 \omega s_3 & -2 \omega^3 s_3 & -2 s_2 & -2 \omega^2 s_2 & -2 \omega^4 s_2 & 2 \omega s_2 & 2 \omega^3 s_2 & 2 s_1 & 2 \omega^2 s_1 & 2 \omega^4 s_1 & -2 \omega s_1 & -2 \omega^3 s_1 \\
 2 s_3 & -2 \omega s_3 & 2 \omega^2 s_3 & -2 \omega^3 s_3 & 2 \omega^4 s_3 & -2 s_2 & 2 \omega s_2 & -2 \omega^2 s_2 & 2 \omega^3 s_2 & -2 \omega^4 s_2 & 2 s_1 & -2 \omega s_1 & 2 \omega^2 s_1 & -2 \omega^3 s_1 & 2 \omega^4 s_1 \\
 2 s_2 & 2 s_2 & 2 s_2 & 2 s_2 & 2 s_2 & 2 s_1 & 2 s_1 & 2 s_1 & 2 s_1 & 2 s_1 & -2 s_3 & -2 s_3 & -2 s_3 & -2 s_3 & -2 s_3 \\
 2 s_2 & 2 \omega^4 s_2 & -2 \omega^3 s_2 & 2 \omega^2 s_2 & -2 \omega s_2 & 2 s_1 & 2 \omega^4 s_1 & -2 \omega^3 s_1 & 2 \omega^2 s_1 & -2 \omega s_1 & -2 s_3 & -2 \omega^4 s_3 & 2 \omega^3 s_3 & -2 \omega^2 s_3 & 2 \omega s_3 \\
 2 s_2 & -2 \omega^3 s_2 & -2 \omega s_2 & 2 \omega^4 s_2 & 2 \omega^2 s_2 & 2 s_1 & -2 \omega^3 s_1 & -2 \omega s_1 & 2 \omega^4 s_1 & 2 \omega^2 s_1 & -2 s_3 & 2 \omega^3 s_3 & 2 \omega s_3 & -2 \omega^4 s_3 & -2 \omega^2 s_3 \\
 2 s_2 & 2 \omega^2 s_2 & 2 \omega^4 s_2 & -2 \omega s_2 & -2 \omega^3 s_2 & 2 s_1 & 2 \omega^2 s_1 & 2 \omega^4 s_1 & -2 \omega s_1 & -2 \omega^3 s_1 & -2 s_3 & -2 \omega^2 s_3 & -2 \omega^4 s_3 & 2 \omega s_3 & 2 \omega^3 s_3 \\
 2 s_2 & -2 \omega s_2 & 2 \omega^2 s_2 & -2 \omega^3 s_2 & 2 \omega^4 s_2 & 2 s_1 & -2 \omega s_1 & 2 \omega^2 s_1 & -2 \omega^3 s_1 & 2 \omega^4 s_1 & -2 s_3 & 2 \omega s_3 & -2 \omega^2 s_3 & 2 \omega^3 s_3 & -2 \omega^4 s_3 \\
\end{psmallmatrix},
\end{equation}
\normalsize
where $s_1=\sin{\frac{\pi}{7}}, s_2=\cos{\frac{\pi}{14}}, s_3=\cos{\frac{3\pi}{14}}, \omega=\sqrt[5]{-1}$ and the primary fields are ordered as
\begin{equation}
    \Vec{\chi}=(\chi_{5, 5}, \chi_{5, 3}, \chi_{5, 1}, \chi_{5, -1}, \chi_{5, -3},
    \chi_{4, 0}, \chi_{4, -2}, \chi_{1, 1}, \chi_{4, 4}, \chi_{4, 2},
    \chi_{2, 0}, \chi_{3, 3}, \chi_{3, 1}, \chi_{3, -1}, \chi_{2, 2}).
\end{equation}
From the argument around Eq. \eqref{eq:Z5-split}, $\chi_{5,5},\chi_{4,0}$ and $\chi_{2,0}$ should be split into two sub-characters.
In other words, the No. $1,6,11$ rows (and columns) in the original $S$ matrix should be split into two.
I arranged them at the No. $16,18,20$ rows (and columns).
The $S$ transformation of sub-characters acquires 3 more characters to be close.
I arranged them at the No. $17,19,21$ rows (and columns).
Finally, we got a $21\times21$ extended $S$ matrix. 
Here is the $1, 6, 11, 16, 17, 18, 19, 20, 21$ rows and columns of $\frac{1}{\sqrt{2}}\hat{S}$
because the matrix is too large
\tiny
\begin{equation}
\begin{psmallmatrix}
\frac{s_1}{\sqrt{70}} & \frac{s_3}{\sqrt{70}} & \frac{s_2}{\sqrt{70}} & \frac{s_1}{\sqrt{70}} & \frac{-2 y^2-2 z^2+1}{2}  & \frac{s_3}{\sqrt{70}} & w z+x y & \frac{s_2}{\sqrt{70}} & x z-w y \\
\frac{s_3}{\sqrt{70}} & -\frac{s_2}{\sqrt{70}} & \frac{s_1}{\sqrt{70}} & \frac{s_3}{\sqrt{70}} & x y-w z & -\frac{s_2}{\sqrt{70}} & \frac{-2 x^2-2 z^2+1}{2} & \frac{s_1}{\sqrt{70}} & w x+y z \\
\frac{s_2}{\sqrt{70}} & \frac{s_1}{\sqrt{70}} & -\frac{s_3}{\sqrt{70}} & \frac{s_2}{\sqrt{70}} & w y+x z & \frac{s_1}{\sqrt{70}} & y z-w x & -\frac{s_3}{\sqrt{70}} & \frac{-2 x^2-2 y^2+1}{2} \\
\frac{s_1}{\sqrt{70}} & \frac{s_3}{\sqrt{70}} & \frac{s_2}{\sqrt{70}} & \frac{s_1}{\sqrt{70}} & \frac{2 y^2+2 z^2-1}{2} & \frac{s_3}{\sqrt{70}} & -w z-x y & \frac{s_2}{\sqrt{70}} & w y-x z \\
\frac{-2 y^2-2 z^2+1}{2} & x y-w z & w y+x z & \frac{2 y^2+2 z^2-1}{2} & 0 & w z-x y & 0 & -w y-x z & 0 \\
\frac{s_3}{\sqrt{70}} & -\frac{s_2}{\sqrt{70}} & \frac{s_1}{\sqrt{70}} & \frac{s_3}{\sqrt{70}} & w z-x y & -\frac{s_2}{\sqrt{70}} & \frac{2 x^2+2 z^2-1}{2} & \frac{s_1}{\sqrt{70}} & -w x-y z \\
w z+x y & \frac{-2 x^2-2 z^2+1}{2} & y z-w x & -w z-x y & 0 & \frac{2 x^2+2 z^2-1}{2} & 0 & w x-y z & 0 \\
\frac{s_2}{\sqrt{70}} & \frac{s_1}{\sqrt{70}} & -\frac{s_3}{\sqrt{70}} & \frac{s_2}{\sqrt{70}} & -w y-x z & \frac{s_1}{\sqrt{70}} & w x-y z & -\frac{s_3}{\sqrt{70}} & \frac{2 x^2+2 y^2-1}{2} \\
x z-w y & w x+y z & \frac{-2 x^2-2 y^2+1}{2} & w y-x z & 0 & -w x-y z & 0 & \frac{2 x^2+2 y^2-1}{2} & 0 \\
\end{psmallmatrix}.
\end{equation}
\normalsize
The $x,y,z$, and $w$ obey $x^2+y^2+z^2+w^2=1$, and they account for the fact that we do not know the exact form of the 3 more characters.
So, there are extra parameters for a $SO(3)$ group.
To fix $x,y,z$ and $w$, we then need to use the restriction that the fusion rule is non-negative integers.
Finally, the solution is
\begin{equation}
    x= 0.814857,y=0.362646,z=0.452212,w=0.
\end{equation}
Then, we can construct the explicit form of the extended $S$ matrix and the fusion rule as Eq. \eqref{eq:Z5-fusionrule}.

\bibliographystyle{JHEP}
\bibliography{references}




\end{document}